%% file: eprint.tex
\documentclass[10pt]{article}
\usepackage{graphicx}

\def\Title#1{\begin{center} {\Large #1 } \end{center}}
\def\Author#1{\begin{center}{ \sc #1} \end{center}}
\def\Address#1{\begin{center}{ \it #1} \end{center}}

\newcommand\pubblock{\rightline{\begin{tabular}{l} Proceedings of the Fifth Annual LHCP Conference\\ \pubnumber\\
         \pubdate  \end{tabular}}}

\newenvironment{Abstract}{\begin{quotation} \begin{center} 
             \large ABSTRACT \end{center}\bigskip 
      \begin{center}\begin{large}}{\end{large}\end{center} \end{quotation}}

\newenvironment{Presented}{\begin{quotation} \begin{center} 
             PRESENTED AT\end{center}\bigskip 
      \begin{center}\begin{large}}{\end{large}\end{center} \end{quotation}}

\input econfmacros.tex

\usepackage{color}

\newcommand\pubnumber{ CMS-CR-2017-238 }

\newcommand\pubdate{\today}

\def\affiliation{
Department of Physics \\
National Taiwan University, Taipei, 10617, Taiwan}

\begin{document}

\large
\begin{titlepage}
\pubblock

\vfill
\Title{  Higgs boson measurements in $WW$, $\tau\tau$ and $\mu\mu$ channels with CMS }
\vfill

\Author{ Arun Kumar \\ On behalf of the CMS Collaboration }
\Address{\affiliation}
\vfill
\begin{Abstract}

This note presents search for the Standard Model Higgs boson in $WW$, $\tau\tau$ and $\mu\mu$ decay channels with the proton-proton collision data collected by the CMS experiment at the LHC. The results have been derived using different amounts of luminosities for different channels.

\end{Abstract}
\vfill

\begin{Presented}
The Fifth Annual Conference\\
 on Large Hadron Collider Physics \\
Shanghai Jiao Tong University, Shanghai, China\\ 
May 15-20, 2017
\end{Presented}
\vfill
\end{titlepage}
\def\thefootnote{\fnsymbol{footnote}}
\setcounter{footnote}{0}
%

\normalsize

\section{Introduction}

In the Standard Model (SM) of particle physics \cite{sm:weinsberg,sm:salam} electroweak symmetry breaking is achieved
via the Brout-Englert-Higgs mechanism, leading to the prediction of the existence of one
physical neutral scalar particle, commonly known as the Higgs boson. 
A particle compatible with such a boson was observed by the ATLAS and CMS experiments at CERN, in the $ZZ$, $\gamma\gamma$ and $WW$ decay channels \cite{higgs:atlas,higgs:cms}, during Run-I of the LHC.
In this note, the latest results on the SM Higgs boson, available at the time of Large Hadron Collider Physics conference in May 2017, from three different decay channels are presented. The data have been collected by the CMS detector~\cite{cms:detector}.

\section{$H\rightarrow \tau\tau$ analysis}
The $H\rightarrow \tau\tau$ analysis has been updated with the full luminosity collected in 2016, corresponding to 35.9~$fb^{-1}$ \cite{htt:run2}.
Four final states are considered depending on the decay of the di-$\tau$ pair: $e\mu$, $e\tau_{h}$, $\mu\tau_{h}$, $\tau_{h}\tau_{h}$, where $\tau_{h}$ denotes the semi-hadronic decay of the $\tau$ lepton.
Semi-hadronically decaying $\tau_{h}$-leptons, are reconstructed with the hadron-plus-strips (HPS) algorithm~\cite{hps:tau}, which is seeded with the anti-kT jets~\cite{jet:antikt}.
The HPS algorithm reconstructs $\tau_{h}$ candidates based on the number of tracks and on the number of ECAL strips with energy deposits, in the one-prong, one-prong + $\pi^{0}$(s), and three-prong decay modes.
The working point used in this analysis has an efficiency of about 60$\%$ for genuine $\tau_{h}$, 
with about 1$\%$ mis-identification rate for quark and gluon jets.
The neutrinos from the $\tau$-lepton decays take away a large fraction of the $\tau$-lepton energy, and this reduces the discriminating
power of the invariant mass of the di-$\tau$ system, in the following denoted as $m_{vis}$.
The SVFIT algorithm combines the missing transverse energy ($E^{miss}_{T}$) with the four-vectors of both 
 $\tau$ candidates, to estimate the value of mass of the parent particle of the di-$\tau$ system, in the following, denoted as $m_{\tau\tau}$.

\subsection{Event categorization and backgrounds}
\begin{figure}[htbp]
\centering
     \includegraphics[width=0.3\textwidth]{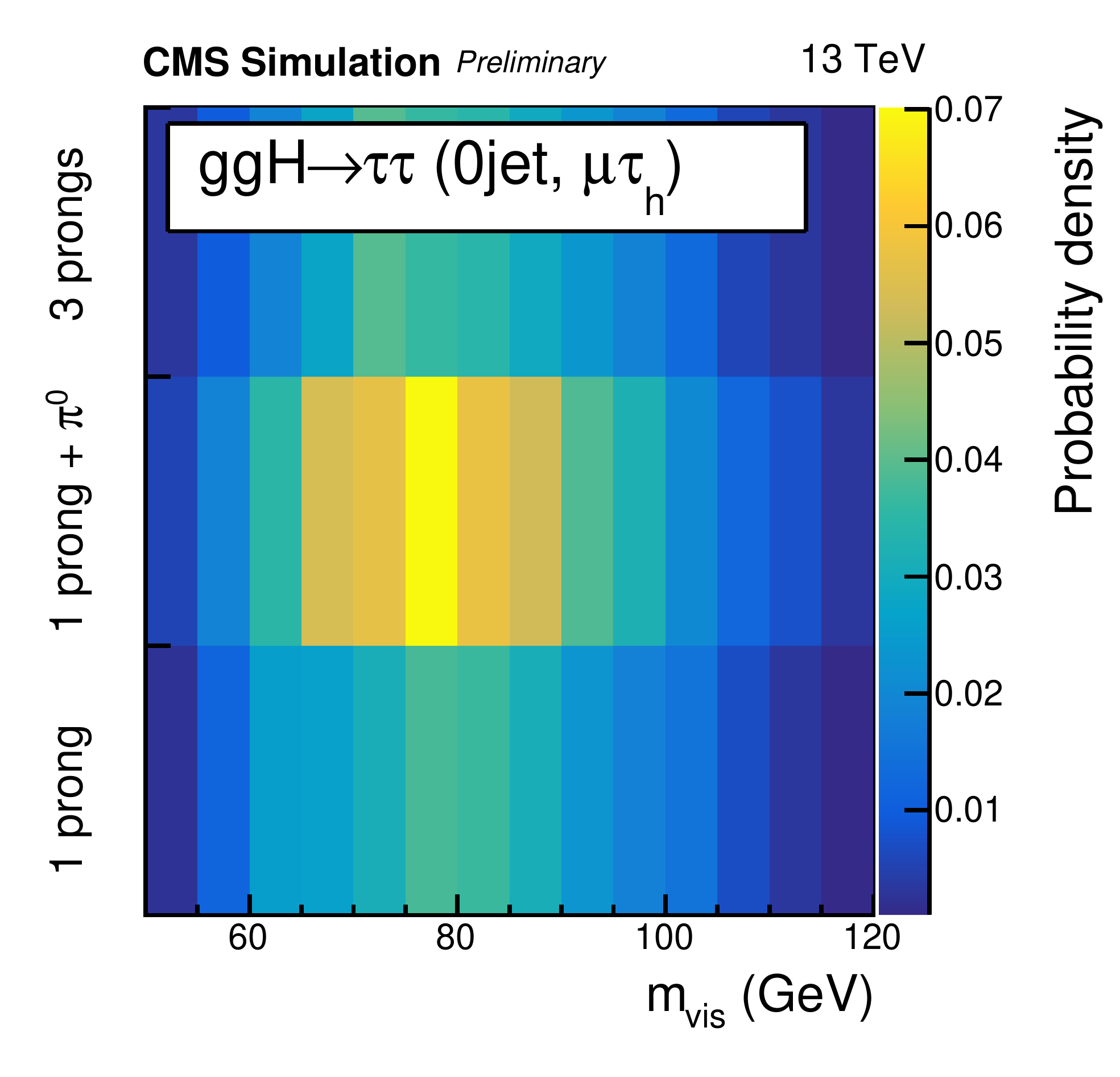}
     \includegraphics[width=0.3\textwidth]{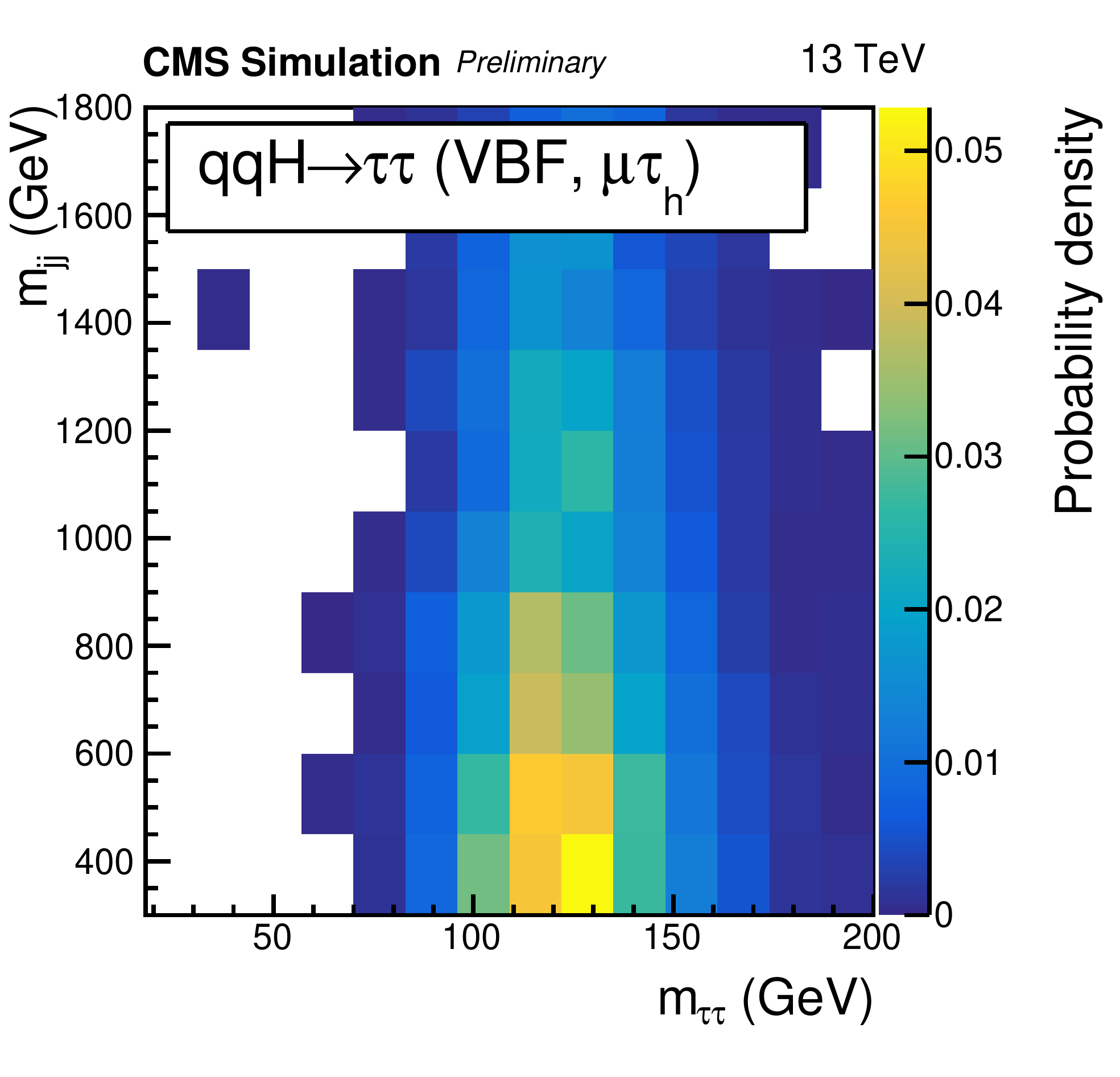}
     \includegraphics[width=0.3\textwidth]{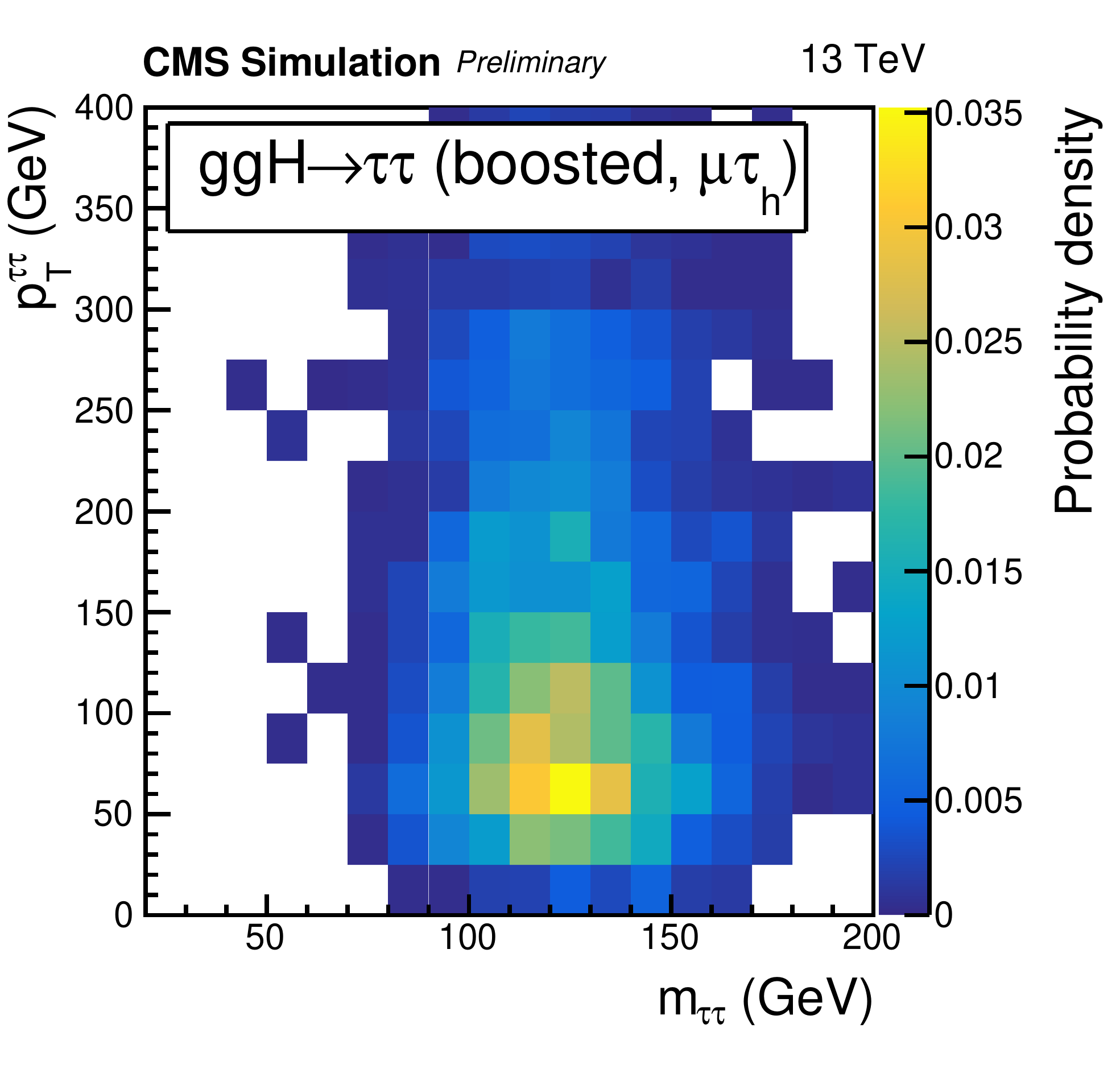}\\
     \includegraphics[width=0.3\textwidth]{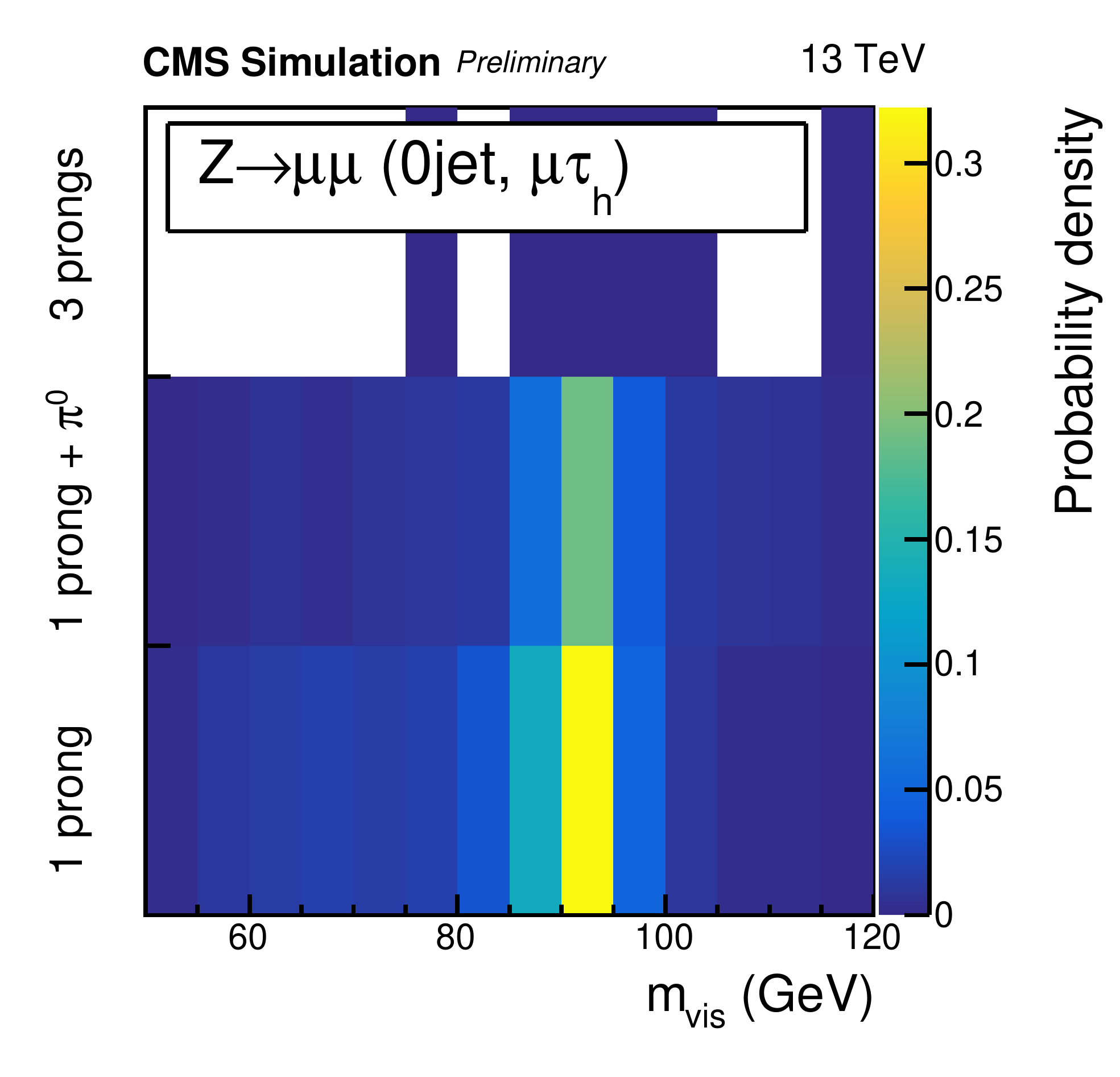}
     \includegraphics[width=0.3\textwidth]{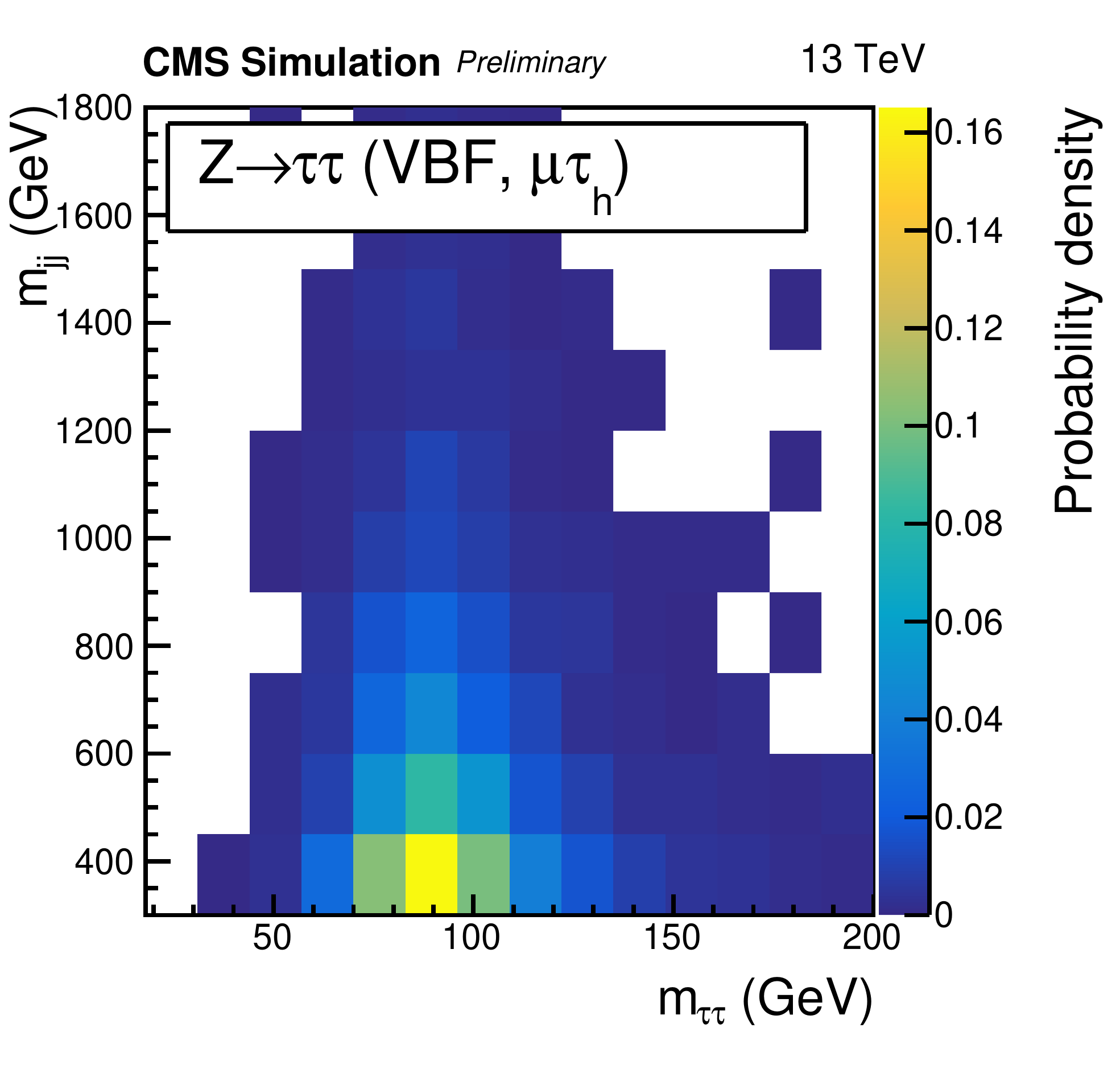}
     \includegraphics[width=0.3\textwidth]{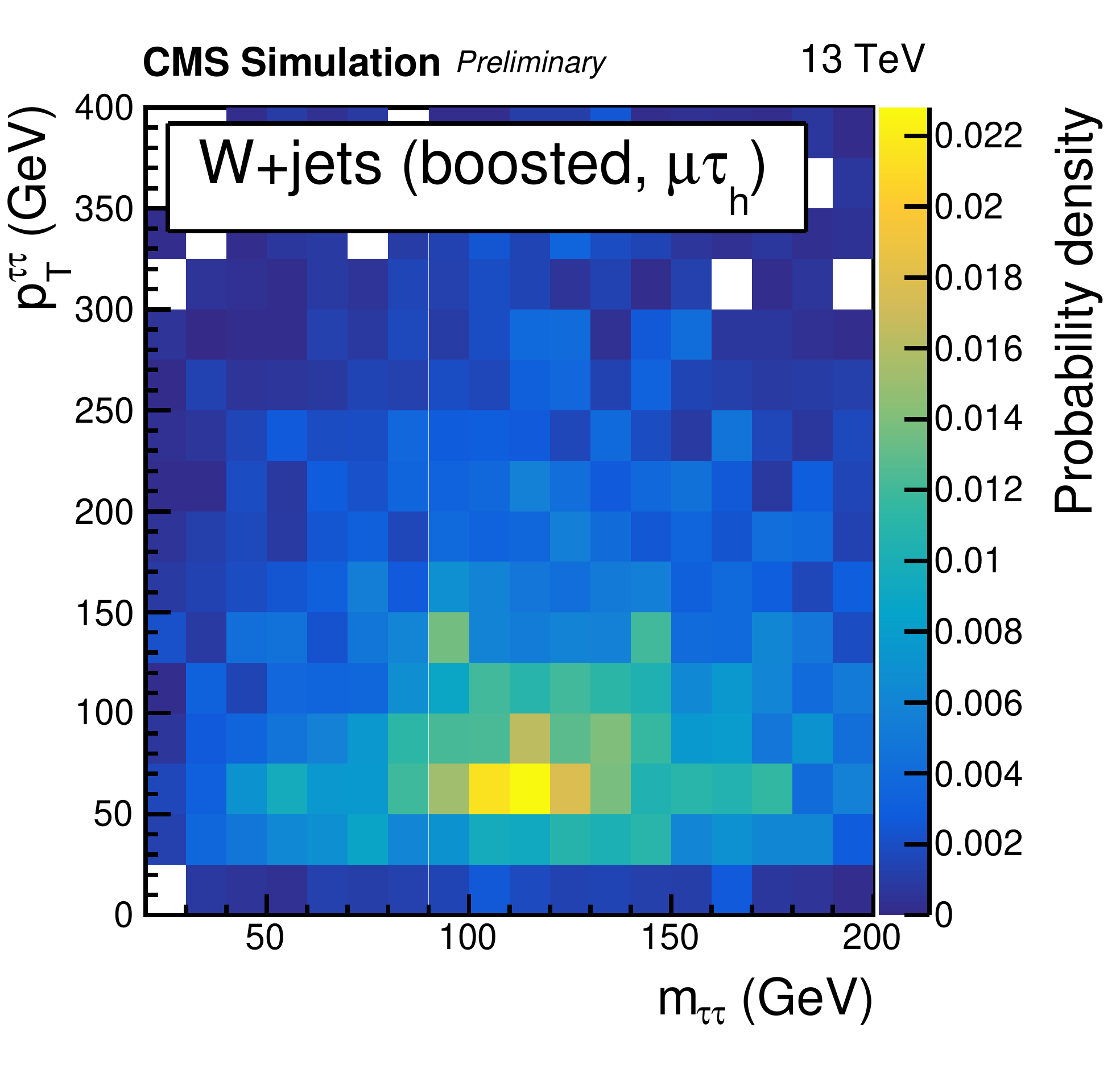}
     \caption{Distributions for the signal (top) and typical background processes (bottom) of the two observables chosen in the 0 jet (left), VBF (centre), and boosted
(right) categories in the $\mu$$\tau_{h}$ final state~\cite{htt:run2}.}
     \label{fig:2Dcategories}
\end{figure}

The events are categorized based on jet mulitplicity into three categories.
\begin{itemize}
\item {\bf 0-jet :} These mostly contains events from gluon-fusion.
Variables which are used to extract the results are $m_{vis}$ and the reconstructed $\tau_{h}$ decay mode (in $\mu\tau_{h}$, e$\tau_{h}$ channel) and the $p_{T}$ of the muon in the e$\mu$ channel. Drell-Yan process is the major background. 2D distributions of the signal and background are shown in Figure \ref{fig:2Dcategories}.
\item {\bf VBF :} This category targets scalar boson events produced via Vector Boson Fusion (VBF).
Events are selected with at least two (exactly two) jets with $p_{T}$ $>$ 30 GeV in the $\mu\tau_{h}$, e$\tau_{h}$ and, $\tau_{h}\tau_{h}$ (e$\mu$) channels.
In the $\mu\tau_{h}$, e$\tau_{h}$ and e$\mu$ channels, the two leading jets are required to have
an invariant mass, $m_{jj}$, larger than 300 GeV. 
The two observables in the VBF category are $m_{\tau\tau}$ and $m_{jj}$.
\item {\bf Boosted :} All the events which do not enter in any of above categories are considered in this category.
The two observables in this category are $m_{\tau\tau}$ and $p_{T}^{\tau\tau}$, where $p_{T}^{\tau\tau}$ is the vectorial sum of the transverse momenta of the reconstructed $\tau$ leptons and $E^{miss}_{T}$.
\end{itemize}
The details of the backgrounds estimation can be found in~\cite{htt:run2}.
Major backgrounds like $DY$, $t\bar{t}$, $W+jets$ are estimated using simulation with corrections taken from data.
QCD background has been estimated using data-driven techniques.

\subsection{Systematics and results}
One of the most dominant sources of uncertainty is the $\tau$ reconstruction.
The identification of $\tau$ leptons contributes around 5$\%$ rate uncertainty.
The trigger efficiency uncertainty per $\tau_{h}$ leg amounts to an additional 5$\%$, which leads to a total of 10$\%$ 
uncertainty for processes estimated from Monte-Carlo (MC) simulations in the $\tau_{h}\tau_{h}$ final state.
Other major exerimental uncertainties include $\tau$ energy scale, jet energy scale, $E^{miss}_{T}$ scale, mis-identification of all leptons and background estimations.
Theoretical uncertainties are considered due to uncertainties in the parton distribution functions (PDF), variations of the renormalization and factorization scales, and uncertainties in the modelling of the underlying event and parton showers (UEPS). These affects the rate and acceptance uncertainties for the signal processes.
The uncertainty in the integrated luminosity amounts to 2.5$\%$ for data collected in 2016; this affects the normalization of processes fully estimated through MC simulations.

\begin{figure}[htbp]
\centering
     \includegraphics[width=0.46\textwidth]{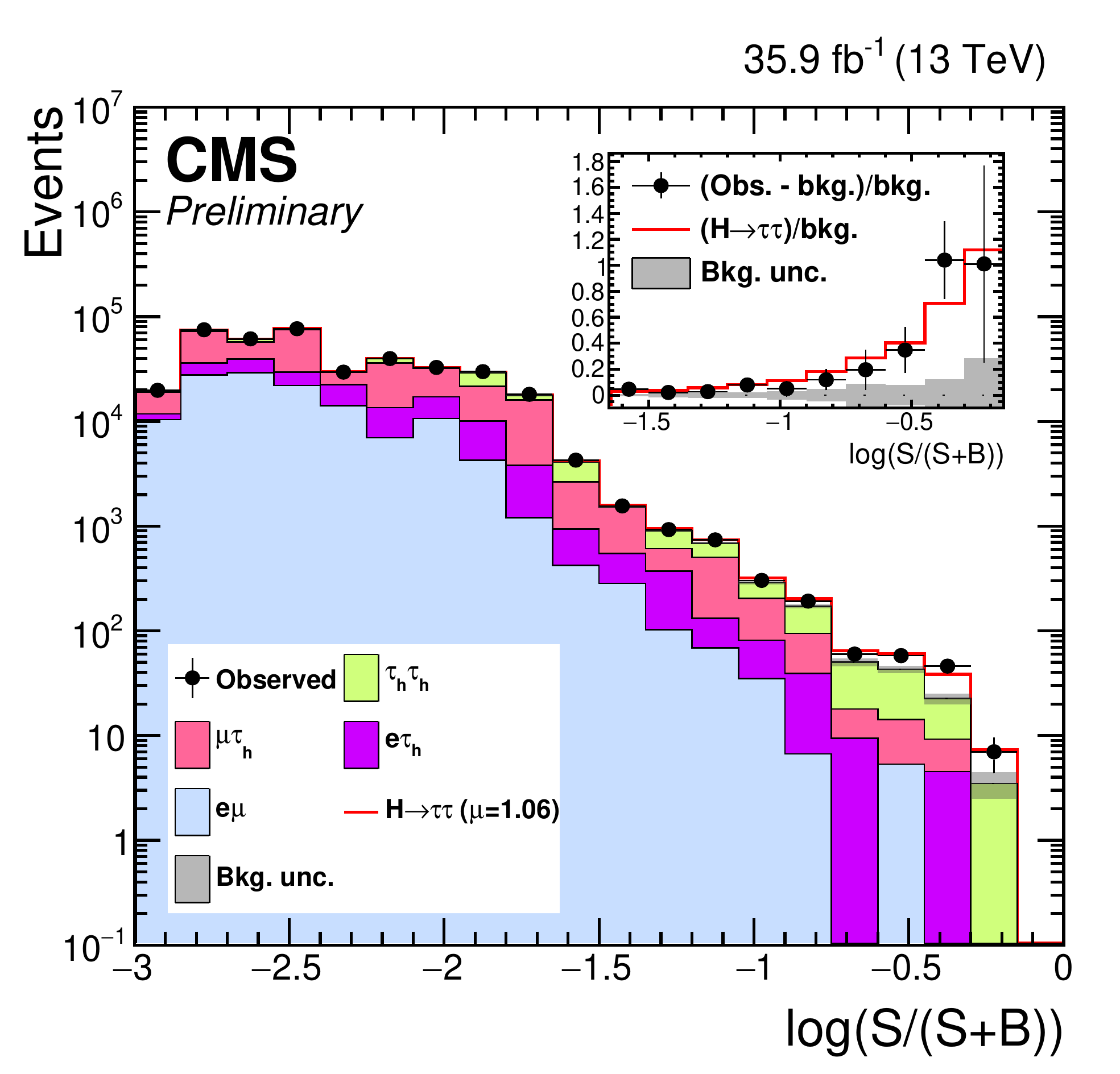}
     \includegraphics[width=0.46\textwidth]{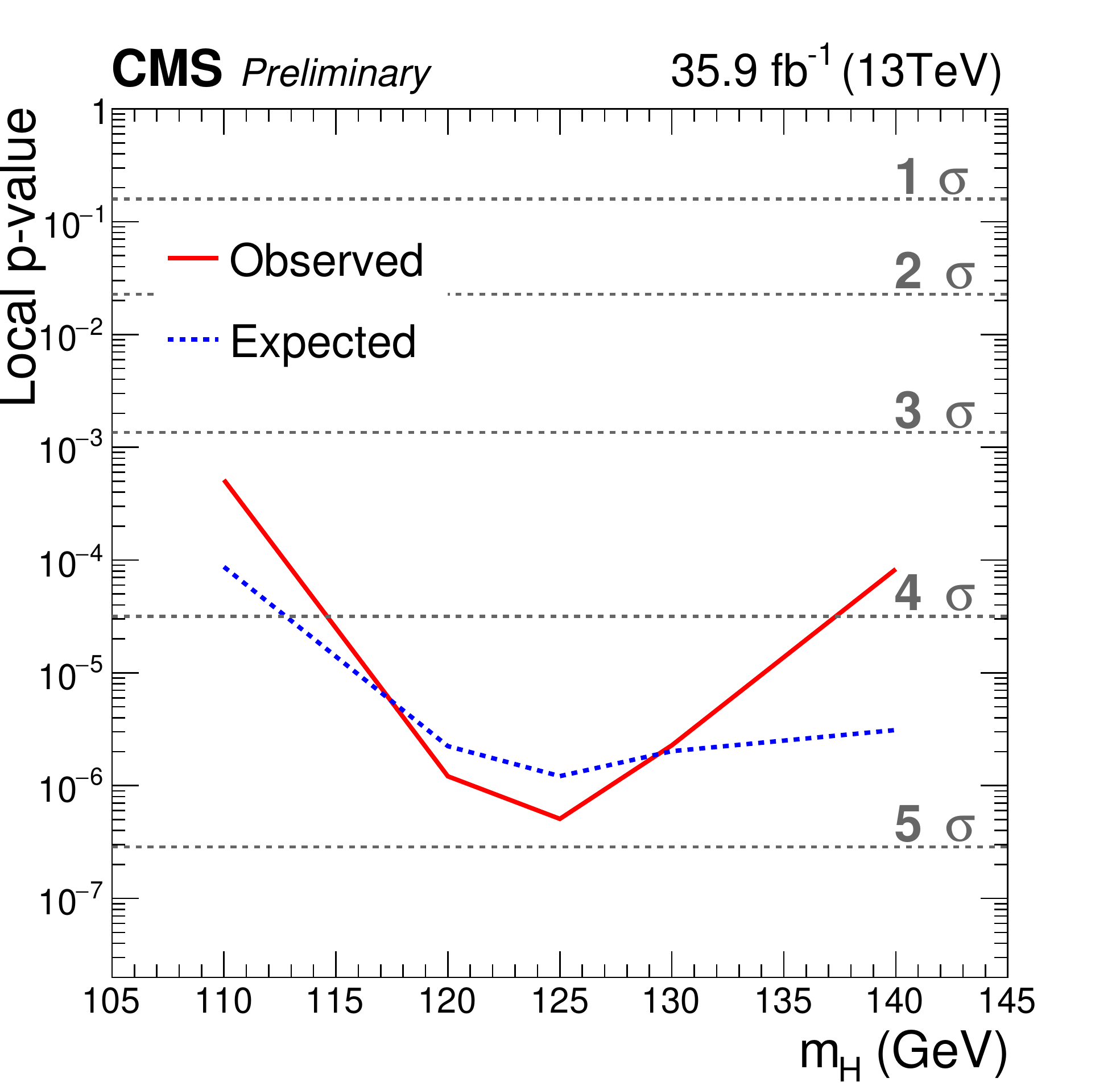}
     \caption{(\textit{left}) Distribution of the decimal logarithm of the ratio between the expected signal and the sum of expected signal and expected background in each bin of the mass distributions used to extract the results, in all signal regions. (\textit{right}) Local p-value and significance as a function of the SM scalar boson mass hypothesis. The observation (black) is compared to the expectation (blue) for a scalar boson with a mass $m_{H}$ = 125 GeV. 
The background includes scalar boson decays to a pair of W bosons, with $m_{H}$ = 125~GeV~\cite{htt:run2}.}
     \label{fig:sbweighted_htt}
\end{figure}

The signal region events are rearranged in a histogram based on the decimal logarithm of the ratio of the signal to signal-plus-background in each bin of the individual distributions used to extract the results. The resulting distribution is shown in Figure~\ref{fig:sbweighted_htt} (\textit{left}). An excess of observed events with respect to the SM background expectation is clearly visible in the most sensitive bins of the analysis.
The channel that contributes the most to these bins is $\tau_{h}\tau_{h}$.
The excess in data is quantified by calculating the corresponding local p-values using a profile-likelihood ratio test statistics. 
As shown in Figure~\ref{fig:sbweighted_htt} (\textit{right}), the observed significance for a SM scalar boson with $m_{H}$ = 125 GeV is 4.9 standard deviations, for an expected significance of 4.7 standard deviations.

\begin{figure}[htb]
  \begin{center}
    \includegraphics[width=0.3\textwidth]{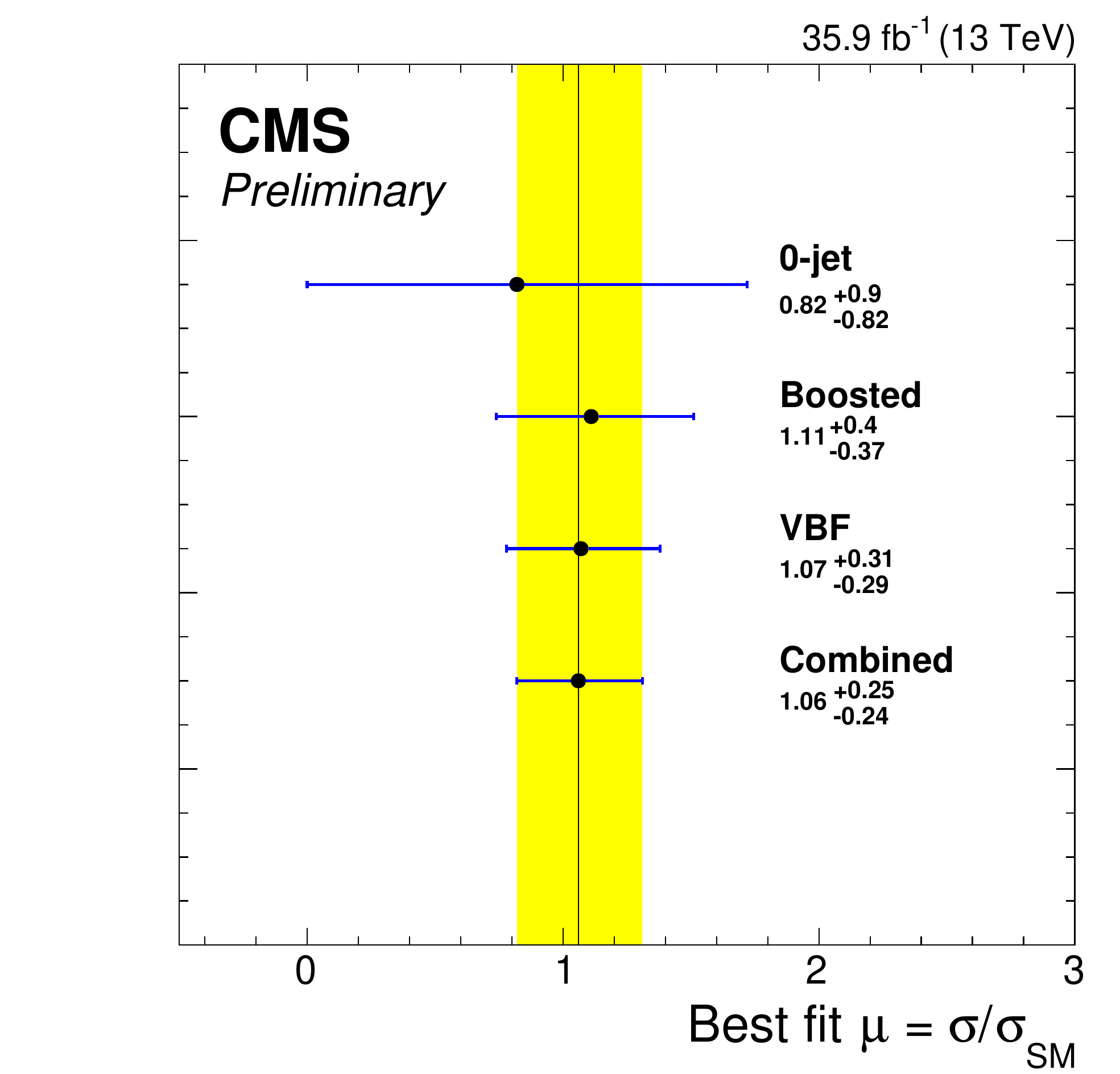}
    \includegraphics[width=0.3\textwidth]{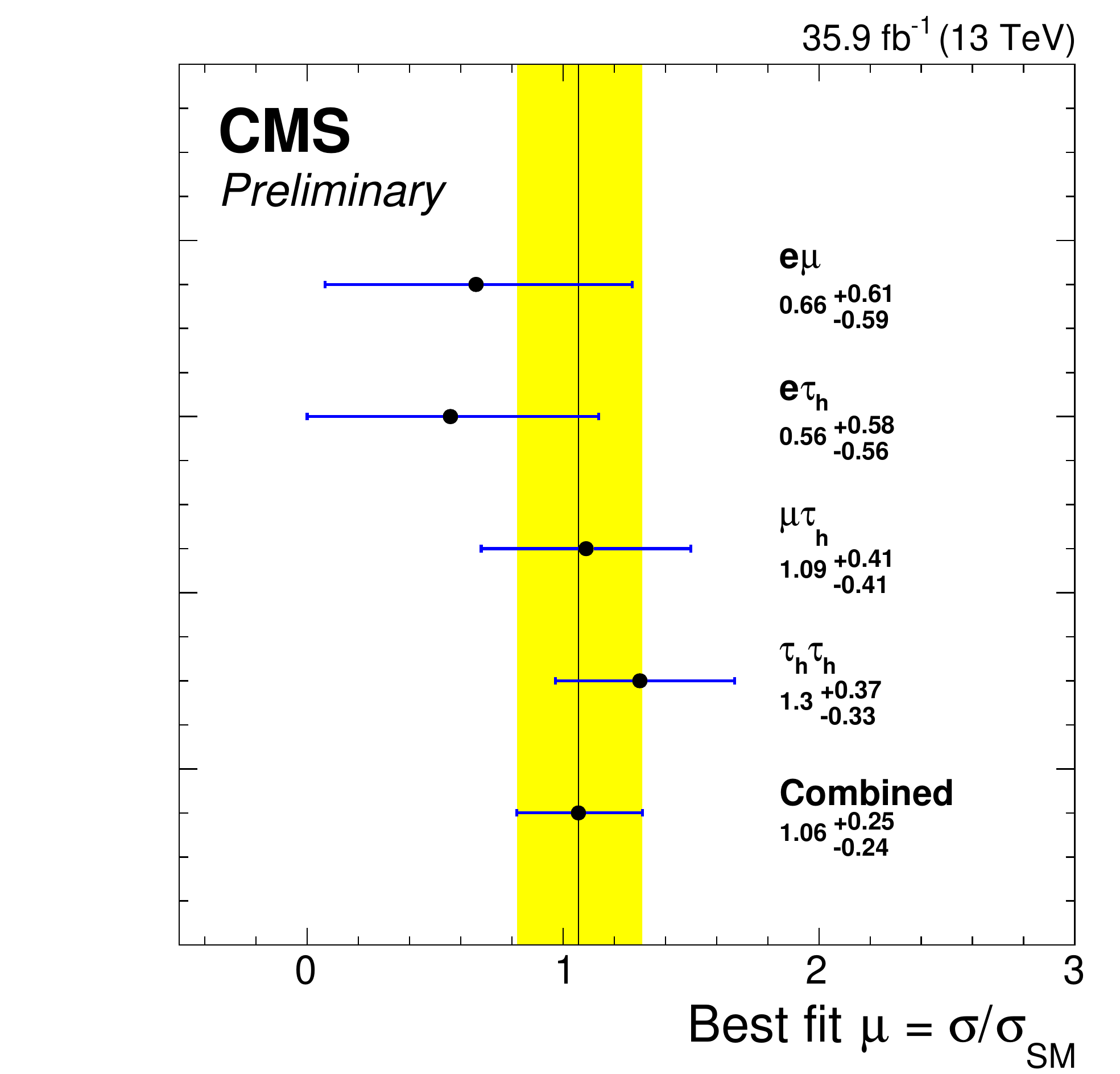}
    \includegraphics[width=0.3\textwidth]{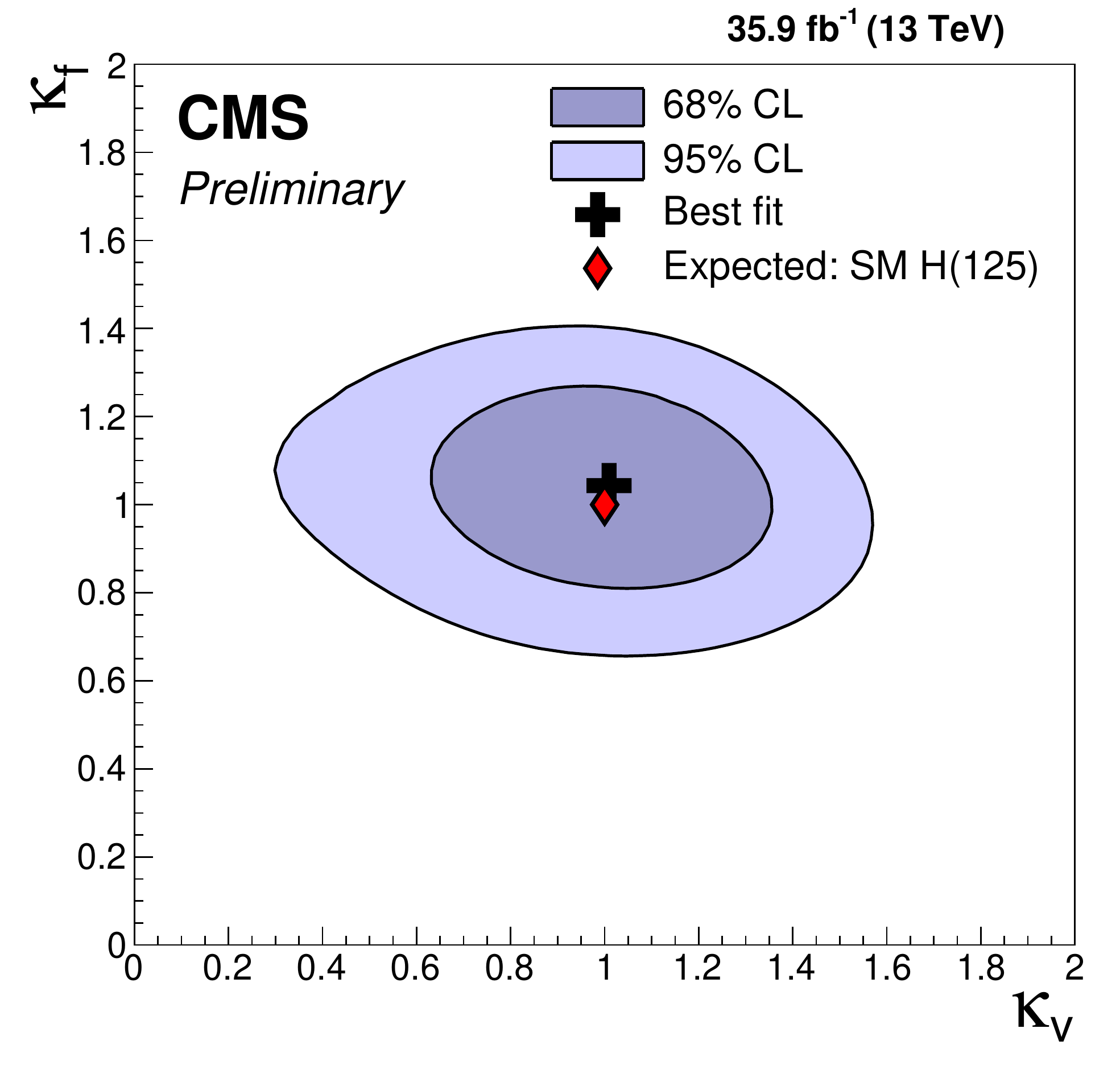}
   \caption{Best-fit signal strength per category (\textit{left}) and channel (\textit{middle}), for $m_{H} = 125$ GeV. The constraints from the global fit are used to extract each of the individual best-fit signal strengths. The combined best-fit signal strength is $\hat{\mu} = 1.06 \pm 0.25$. (\textit{right}) Scan of the negative log-likelihood difference as a function of $\kappa_V$ and $\kappa_f$, for $m_{H}$ = 125~GeV. 
All nuisance parameters are profiled for each point. For this scan, the $pp\rightarrow H(125 GeV)\rightarrow WW$ contribution is treated as a signal process~\cite{htt:run2}.}
    \label{fig:muvalue}
  \end{center}
\end{figure}

The corresponding best-fit value for the signal strength $\mu$ is $\hat{\mu}$ = 1.06 $\pm$ 0.25 at $m_{H}$ = 125 GeV.
The uncertainty on the best-fit signal strength can be decomposed into four components: theoretical
uncertainties, bin-by-bin statistical uncertainties on the backgrounds, other systematic uncertainties, and 
statistical uncertainty. In that format, the best-fit signal strength is $\mu$ = $1.06^{+0.11}_{-0.09}
(th.)~^{+0.12}_{-0.12} (bbb.)~^{+0.13}_{-0.12} (syst.)~^{+0.15}_{-0.15}(stat.)$. 
The individual best-fit signal strengths per channel and per category are given in Figure~\ref{fig:muvalue}; they demonstrate the channel and category-wise consistency of the observation with the SM scalar boson hypothesis.

A likelihood scan is performed for $m_{H}=125$ GeV in the ($\kappa_V$,$\kappa_f$) parameter space, where $\kappa_V$ and $\kappa_f$ quantify, respectively, the ratio between the measured and the SM value for the couplings of the scalar boson to vector bosons and fermions. For this scan, scalar boson decaying to a pair of W bosons are considered as a part of the signal (for the signal strength instead, they are considered as background). All nuisance parameters are profiled for each point of the scan. As shown in Figure~\ref{fig:muvalue}, the observed likelihood contour is consistent with the SM expectation of $\kappa_V$ and $\kappa_f$ equal to unity.

\section{$H\rightarrow WW$ analysis}

This decay mode has been studied in Run-I using the full luminosity collected \cite{hww:run1}, taking into account different production modes.
The update presented here shows the first results from the $H\rightarrow WW$ analysis at $\sqrt {s}$ = 13 TeV using a total integrated luminosity of 2.3 $fb^{-1}$ \cite{hww:run2}. 
Only the gluon-fusion production mode has been considered in this update.
Final states containing opposite flavour leptons are considered in this analysis along with missing transverse energy. Events are further categorized based on the presence of jets. Final states with zero or one jet are considered.   
The dominant irreducible background is $WW$ production, which can be discriminated from the signal by means of the lepton kinematics.
Top-related processes ($t\bar{t}$, single-top) are the second dominant background.
Other processes such as Drell-Yan, $W$+jets and other electroweak productions are minor contributions.

\subsection{Event selection}
The analysis strategy has been designed taking into account events with 0-jet and 1-jet only: in the gluon-fusion production mode, the extra jets come from initial or final state radiation.
Only mixed flavour leptonic final states $e^{+}\bar{\nu_{e}}\mu^{-}\nu_{\mu}$ or $e^{-}\nu_{e}\mu^{+}\bar{\nu_{\mu}}$ are considered in this early update, to suppress the large DY background in the same flavour final states.
The indirect contribution from $\tau$ leptons decaying leptonically is also included.
Events are selected using single or double lepton triggers, with a condition of exactly one electron and muon with opposite charge with a minimum $p_{T}$ of 10 (13)~GeV for the muon (electron). 
One of the two leptons should have $p_{T}$ greater than 20~GeV.
Both leptons are required to be well identified and isolated to reject fake leptons.
Events with additional identified and isolated leptons with $p_{T}$ $>$ 10 GeV are rejected.
The low dilepton invariant mass ($m_{ll}$) region dominated by QCD production of leptons is not considered in the analysis with $m_{ll} >$ 12~GeV.
A requirement on the transverse missing energy at 20~GeV is imposed to mainly suppress DY.
The DY background is further reduced by requesting the di-lepton transverse momentum ($p_{T}^{ll}$) to be higher than 30 GeV.  
These selection criteria also reduce contributions from $H\rightarrow \tau\tau$ in the present analysis.
To reduce the contamination of single top and top pair production, it is required to have no jets with $p_{T}~>~20$ GeV, identified as $b$-quark jets.

\subsection{Analysis strategy and background estimation}
To further enhance the sensitivity to the SM Higgs boson signal, events are categorized based on the number of jets with $p_{T}$ above 30 GeV. 
The zero jet category is dominated by the non-resonant $WW$ background while in the one jet category the contributions from top backgrounds are of similar nature to that of non resonant $WW$ process.
Higher jet multiplicity categories that would be more sensitive to Higgs production mechanisms other than gluon fusion are not considered in the present analysis. 
To disentangle another important background, $W+jets$, where one jet fakes an isolated lepton, the 0-jet and 1-jet categories are further split according to the lepton flavour: $e\mu$ and $\mu e$, where the first lepton is the one with higher transverse momentum. 
The improvement in terms of significance is about 10$\%$ taking advantage of the different probabilities for a jet to mimic an electron or a muon.
The final discriminant to extract the signal and perform statistical analysis is a 2-dimensional variable based on $m_{ll}$ and the transverse mass of the $WW$ system ($m_{T}^{H}$) = $\sqrt{2p_{T}^{ll}E_{T}^{miss}(1-cos\Delta\phi(ll,\vec{E_{T}}^{miss}))}$, where $\Delta\phi(ll,\vec{E_{T}}^{miss})$ is the azimuthal angle between the dilepton momentum and $E_{T}^{miss}$.
\subsubsection{Background Estimation}
\begin{itemize}
\item Jet-induced backgrounds are estimated using data-driven methods in which a control sample is defined with events, where one lepton passes the standard lepton identification and isolation criteria and another lepton candidate fails the criteria but passes a looser selection.
The probability for a jet satisfying this looser selection to pass the standard one is estimated directly from data in an independent control sample dominated by events with non-prompt leptons from multijet processes. 
This probability is parametrized as a function of $p_{T}$ and $\eta$ of the leptons and applied to the control sample to reproduce the kinematics of the expected jet-induced background in the signal region.
The jet-induced background estimation has relatively large uncertainties due to its dependence on the sample composition, and the limited size of the control sample.
\item Top-related backgrounds are estimated in two steps. 
A scale factor is computed to account for the different efficiencies in data and simulation, such as the discrepancies related to different b-tagging efficiencies and mistag rates. 
This correction is applied by reweighting all the simulated samples with a weight per event which depends on the number of jets, on their flavour, and on the jet kinematics. 
This reweighting allows to correct for shape differences between data and simulation but does not assure that the overall normalization of the simulated samples is correct. 
A top quark enriched control region is defined in data and used to normalize the simulation to the data in that region. 
This is then used to estimate the top background contribution to the signal region. 
\item Other backgrounds like DY and $W\gamma^{*}$ are estimated using simulation, with normalization taken from control samples made with data.
\end{itemize}

\subsection {Results}
All experimental sources of uncertainty, except luminosity, are treated both as normalization and shape uncertainties. 
For backgrounds with a data driven normalization estimation, only the shape uncertainty is considered. 
All the experimental uncertainties are correlated among the signal and background processes in all the categories.
More details of the systematics can be found here~\cite{hww:run2}.


Combining the four categories, the observed (expected) significance is 0.7$\sigma$ (2.0~$\sigma$)
for a SM Higgs boson with a mass of 125 GeV. The corresponding best fit signal strength, $\sigma/\sigma_{SM}$,
which is the ratio of the measured $H\rightarrow$WW$\rightarrow~e\nu\mu\nu$ signal yield to the expectation for
the SM Higgs boson is $0.3 \pm 0.5$.


\section{$H\rightarrow \mu\mu$ analysis}

For a Higgs boson mass, $m_{H}$, of 125 GeV, the SM prediction for the Higgs to $\mu^{+}\mu^{-}$ branching fraction,
is among the smallest accessible at the LHC, $2.2 \times 10^{-4}$.
Despite the small branching fraction, this channel has the cleanest experimental signature among the Higgs fermonic decay channels.
This analysis may be used to test if the coupling of the new boson to leptons is flavour-universal or proportional to the lepton
mass, as predicted by the SM. 
In addition, a measurement of the $H\rightarrow \mu\mu$ decay probes the Yukawa coupling of the Higgs boson to second-generation fermions, an important input
in understanding the mechanism of electroweak symmetry breaking in the SM.
The most recent analysis is still the one based on Run-I dataset~\cite{hmumu:run1}.
These data correspond to integrated luminosities of 5.0 fb$^{-1}$ at a centre of mass energy of 7~TeV, and 19.7 fb$^{-1}$ at 8~TeV.

\subsection{Event selection and analysis strategy}
Events are selected using the single muon trigger which requires an isolated muon with $p_{T}~>$ 24~GeV and $|\eta|~\leq$ 2.1.
Tight identification and isolation criteria~\cite{hmumu:muonid} have been applied to select good quality muons.
Every muon is required to have an impact parameter with respect to the primary vertex smaller than 5 and 2~mm in the longitudinal and transverse directions, respectively. 
They must also have $p_{T}~>$15 GeV and $|\eta|~\leq$ 2.1.
Muons which triggered the event must have $p_{T}~>$ 25 GeV and after that all combinations of opposite sign di-muon pairs are made for the invariant mass distribution analysis.
Each di-muon pair is effectively treated as a separate event.
After selecting events with a pair of isolated opposite-sign muons, events are categorized according to the properties of jets. 
Jets with $p_{T}~>$15 GeV and $|\eta|~\leq$ 4.7 are considered.

Di-muon events are categorized as 2-jet category and 0, 1-jet category.
The 2-jet category requires at least two jets, with $p_{T}>40$ GeV for
the leading jet and $p_{T}>30$ GeV for the subleading jet. 
A requirement on $E_{T}^{miss}<40$ GeV is imposed in 2-jet events to reduce the number of $t\bar{t}$ events.
The rest of the events are kept in the 0, 1-jet category where the signal is dominated by gluon-fusion.

The 2-jet category is further split into VBF~Tight,
gluon-fusion~Tight, and Loose subcategories.
The VBF Tight category is defined to have $M_{jj}>650$ GeV and $|{\Delta\eta_{jj}}|>3.5$, where $|{\Delta\eta_{jj}}|$
is the absolute value of the difference in pseudorapidity between the two leading jets.
For a SM Higgs boson with $m_{H}=125$ GeV, 79\% of the signal events in this category are from VBF production.

The gluon-fusion~Tight category captures these events by requiring
the di-muon transverse momentum to be greater than 50 GeV and $M_{jj}>250$ GeV.
In the 0, 1-jet category,
events are split into two subcategories based on the value of di-muon $p_{T}$. 
The S/B ratio is further improved by categorizing events based on the di-muon invariant mass
resolution based on $\eta$ of muons. More details can be found in \cite{hmumu:run1}.

\subsection{Results}
The di-muon invariant mass distribution is fit for signal and background shapes using parametrized functions.
The signal shape is a double gaussian whose parameters were obtained from simulation.
The background is dominated by the $DY\rightarrow \mu\mu$ process, and is modeled by a continuous function, $f(M_{\mu\mu})$, which is the sum
of a Breit-Wigner function and a 1/$M_{\mu\mu}^{2}$ term, to model the $Z$-boson and photon contributions, both multiplied
by an exponential function to approximate the effect of the PDF on the $M_{\mu\mu}$ distribution.
This function is described in \cite{hmumu:run1}. The parameters of this function are obtained from a fit to the data.

All the systematic uncertainties are divided into two main parts: shape uncertainties which affects the shape of the di-muon invariant mass spectrum
and rate uncertainties which affects the signal yield in each category.
Muon momentum scale and resolution are the only major sources of shape uncertainties and they affect the width of the signal peak by 3$\%$.
Rate uncertainties in the signal yield are evaluated separately for each Higgs boson production process and each centre-of-mass energy.
The observed 95$\%$ CL upper limits on the signal strength at 125 GeV are 22.4 using the 7~TeV data and 7.0 using the 8~TeV data.
The corresponding background-only expected limits are
$16.6^{+7.3}_{-4.9}$ using the 7 TeV data and
$7.2^{+3.2}_{-2.1}$ using the 8 TeV data.
Accordingly, the combined observed limit for 7 and 8 TeV is 7.4,
 while the background-only expected limit is $6.5^{+2.8}_{-1.9}$.
This corresponds to an observed upper limit on $\mathcal{B}(H\rightarrow \mu\mu)$ of 0.0016, assuming the SM cross section.
The best fit value of the signal strength for a Higgs boson mass of 125 GeV is $0.8^{+3.5}_{-3.4}$.




\end{document}

%% file: econfmacros.tex
\def\beq{\begin{equation}}
\def\eeq#1{\label{#1}\end{equation}}
\def\eeqn{\end{equation}}

\def\beqa{\begin{eqnarray}}
\def\eeqa#1{\label{#1}\end{eqnarray}}
\def\eeqan{\end{eqnarray}}

\let\bar=\overbar

\def\Dslash{\not{\hbox{\kern-4pt $D$}}}
\def\dslash{\not{\hbox{\kern-2pt $\del$}}}

\def\msb{{\bar{\ssstyle M \kern -1pt S}}}

